\documentclass[a4paper,twoside]{article}

\usepackage{epsfig}
\usepackage{subcaption}
\usepackage{graphicx}
\usepackage{comment}
\usepackage{tikz}
\usepackage{amsfonts}
\usepackage{float}
\usepackage{listings}
\usepackage{url}
\usepackage{algorithm}
\usepackage{algpseudocode}
\usetikzlibrary{shapes,backgrounds,calc,arrows}
\usepackage{microtype}
\usepackage{calc}
\usepackage{amssymb}
\usepackage{amstext}
\usepackage{amsmath}
\usepackage{amsthm}
\usepackage{multicol}
\usepackage{pslatex}
\usepackage{apalike}
\usepackage[bottom]{footmisc}
\usepackage{SCITEPRESS}     % Please add other packages that you may need BEFORE the SCITEPRESS.sty package.
\usepackage{atbegshi}% http://ctan.org/pkg/atbegshi
\AtBeginDocument{\AtBeginShipoutNext{\AtBeginShipoutDiscard}}

\begin{document}

\title{Visualizing, Analyzing and Constructing L-system from Arborized 3D Model Using a Web Application}

\author{\authorname{Nick van Nielen\sup{1}, Fons Verbeek\sup{1}, Lu Cao\sup{1}}
\affiliation{\sup{1}Leiden Insisute of Advanced Computer Science, Leiden University, Leiden, The Netherlands}
\email{l.cao@liacs.leidenuniv.nl}
}


\abstract{In biology, arborized structures are well represented and typically complex for visualization and analysis. In order to have a profound understanding of the topology of arborized 3D biological model, higher level abstraction is needed. We aim at constructing an abstraction of arborized 3D biological model to an L-system that provides a generalized formalization in a grammar to represent complex structures. The focus of this paper is to combine 3D visualization, analysis and L-system abstraction into a single web application. We designed a front-end user interface and a back-end. In the front-end, we used A-Frame and defined algorithms to generate and visualize L-systems. In the back-end, we utilized the Vascular Modelling Toolkit's (VMTK) centerline analysis methods to extract important features from the arborized 3D models, which can be applied to L-system generation. In addition, two 3D biological models: lactiferous duct and artery are used as two case studies to verify the functionality of this web application. In conclusion, our web application is able to visualize, analyse and create L-system abstractions of arborized 3D models. This in turn provides workflow-improving benefits, easy accessibility and extensibility.}

\onecolumn \maketitle \normalsize \setcounter{footnote}{0} \vfill

\section{\uppercase{Introduction}}
\label{sec:introduction}
Analysis and visualization of 3D models are unequivocally important in the biological studies. They have been used to study physiological aspects, such as simulating the flow and particle transport of airways~\cite{fluiddynamics}. This 3D model simulation made it possible to obtain particle deposition patterns and flow characteristics that could not be obtained experimentally. 3D models have also been used to analyse the complex anatomy of the canal network in cortical bone~\cite{tomography} to get a better understanding of the mechanical properties. Furthermore, 3D models are used to analyze the phenotypical changes of the lactiferous duct of newborn mice under the exposure of endocrine disruptors~\cite{thesis}. In particular, these articles discuss the usage of 3D models with a tree-like branching structure (or arborized 3D models). These types of structures can be found in plants, but also in animals, such as lactiferous duct or vascular system. Typically, arborized structures are complex due to their high branching order and visibly indistinguishable individual branches. To realize appropriate visualization and interpretation of these complex structures, model abstraction with minimal loss of features is needed. Theoretical computer science methods can be used to extract important features and represent these features in an abstracted model. A well-known system that preserves most of the anatomical features of arborized structures is Lindenmayer systems (L-systems)~\cite{prusinkiewicz1990algorithmic}. We will use L-systems in our research to describe and build abstractions of arborized 3D models.\\
Many different applications already exist to visualize and analyse (arborized) biological 3D models. A popular platform is 3D Slicer - a platform for medical imaging which provides image registration, interactive visualization, model-based analysis, and more advanced functionality~\cite{Kikinis2014}. This can be used together with tools such as the Vascular Modelling Toolkit (VMTK) to analyse vascular-like or arborized structures~\cite{VMTK}. When it comes to the abstraction of models, specifically for generating L-systems, L-py can be used for complex grammar~\cite{boudon:hal-00831780}. More general 3D model visualization, processing and editing tools can also be found, for example Blender~\cite{foundationblender} and MeshLab~\cite{meshlab}. MeshLabJS~\cite{meshlabjs} is a web-based version of MeshLab that uses the JavaScript library Three.js~\cite{threejs} as a rendering engine. ~\cite{Sawicki2013} has shown that web applications can be a well-structured and convenient way of visualizing 3D models. While having clear limitations in handling complex models due to computational and spatial complexities, web applications can improve accessibility and portability from different devices and browsers. Furthermore, web applications do not have to be downloaded or maintained by the end-user and the 3D models can be stored on and loaded from a database, improving the ease of use and allowing for the possibility to share models between multiple users.\\
Both system-based and web-based applications can realize parts of the process needed to visualize, analyse and create abstractions of arborized 3D models. However, not many applications exist that combine these parts which are typically related, especially in a web-based format. To the best of our knowledge, there does not seem to exist such a web-based application that is capable of visualizing, analyzing and creating L-system abstractions of arborized 3D models. \\
In this paper, A-frame was chosen to construct the web-based application because of its rendering performance, cross-compatibility and HTML Entity-Component system. A-frame is a JavaScript web framework for creating (complex) 3D scenes with additional VR functionality. It is built on top of the rendering engine Three.js~\cite{aframe}. Concerning L-system generation, the JavaScript library lindenmayer.js was selected to interpret and generate L-system grammar in an object-oriented fashion~\cite{lindenmayerjs}. In addition, the waterfall development method was adopted to ensure good quality, performance, programming and design~\cite{balaji2012waterfall}.

\section{\uppercase{Method}}\label{methodology}
To visualize and analyse 3D models as well as generate L-systems, we need to design the desired functionality. We will apply design patterns, discuss the design choices and sketch the functionality of our web application using A-Frame in section \ref{design}. Furthermore, in section \ref{modelanalysis} we discuss how the model analysis is implemented with VMTK, python and the user interface of the web application. Lastly, we reason how we combine Three.js and Lindenmayer.js to generate and visualize L-systems in section \ref{lsystemmethodology}.

\subsection{Design}\label{design}
For the design of web-based applications, a modelling method is needed to cover all the requirements. This can help visualize and document the components and corresponding functionality. UML diagrams have been shown to be a powerful modelling method for covering all required components of web applications~\cite{koch2002expressive}. The UML diagrams of the front-end and back-end layout of our web application are available in \url{https://surfdrive.surf.nl/files/index.php/s/8weiJINAJTbplVs}. 

\subsubsection{Front-end and back-end}\label{frontend}
For 3D model visualization and manipulation, a front-end is needed for the user input, display and storing between static web page states. All these functionalities are available on a single page called \verb|index.html|. Furthermore, \verb|index.html| is the view node of the Model-View-Controller (MVC) programming paradigm ~\cite{krasner1988description}, which is used to display the user interface and is a part of the interface component. Besides the view node, the states between static pages are stored in the cookie object data structure altered by the model node in \verb|Settings.js| and updated in the view node. Modifications to this data structure are made by the controller node \verb|UserInterface.js|, which reacts to the user input in the view. Some user input does not change the model and is directly updated in the view by the controller.\\

For the 3D model rendering and visualization, two components are used which contain A-Frame with Three.js rendering. The first component is the main A-Frame. In this component, the main object or L-system view is stored and displayed using the A-Frame web framework inside \verb|aframe.html|. The second component is the preview A-Frames. As the name suggests, this is where all the preview nodes of the L-system, model and A-Frame settings are shown. The L-system generation as well as the 3D model loading from I/O are initiated by the controller using the preview event handler.\\

Tasks in the user interface, that require computations on the server or I/O operations, are realized in back-end. The back-end uses python to run the development server \verb|devServer.py| using flask, a micro web application framework for simple scalability~\cite{grinberg2018flask}. For the deployment server, Gunicorn~\cite{gunicorn2017http} and Nginx~\cite{reese2008nginx} are used with flask as an HTTP interface and web server with multi-threading.\\

\subsection{Model Analysis}\label{modelanalysis}
\begin{figure}[H]
\begin{center}
\includegraphics[height=3.5cm]{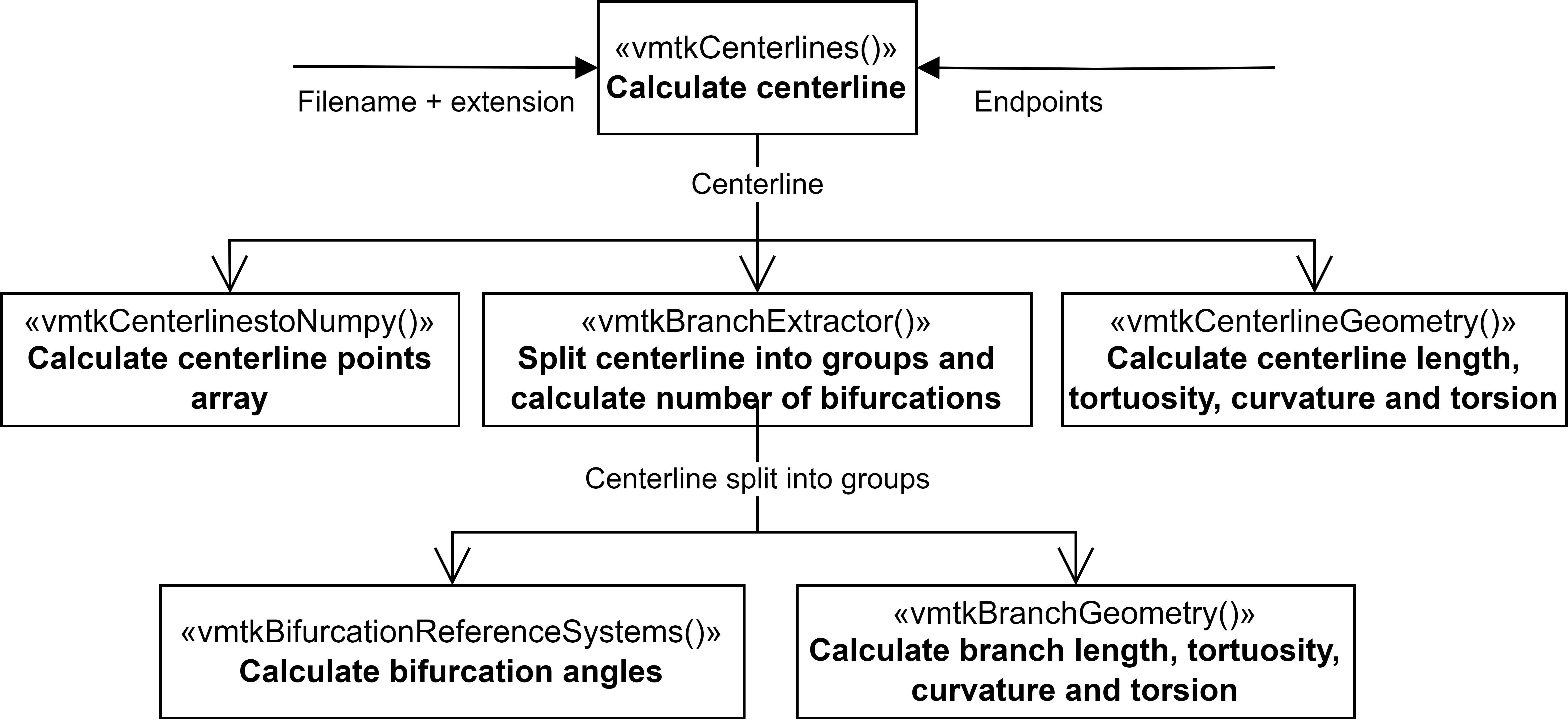}
\end{center}
\caption{Diagram of the analysis process of an arborized 3D model.}
\label{diagramanalysis}
\end{figure}

In the procedure of model analysis, a simplification of the arborized 3D model is needed while preserving the important topological features. The method of center-line extraction \cite{verbeek2020systems} was applied to preserve a one-dimensional representation of the original skeletal structure in 3D. Many types of center-line extraction methods were considered, but the 3D geometric analysis tool  VMTK\cite{VMTK} was chosen. Using the well-defined back-end structure, VMTK tool that runs on python are deployed in the request handler node. During the analysis, endpoints on the model are selected in the front-end view node. The endpoints and 3D model name are then send with an analysis request to the request handler by the controller node. In the request handler, three steps take place: verification, instance creation and the analysis.\\
For verification, the request parameters are verified by checking the 3D model filename and type. If either the filename or file type are incorrect, the model cannot be found in the objects folder and no analysis will be possible. Thus, an error will occur. For the second step, an analysis instance is created on a new thread to prevent request blocking and fatal errors from occurring on the development or deployment thread. The analysis process and its structure can be explained by the diagram in Figure \ref{diagramanalysis}. Endpoints and the 3D model are used to compute the model centerline. Subsequently, it is used to calculate the centerline mesh points. For the final data extractions, the centerline is used to calculate branch length, tortuously, torsion, curvature and bifurcation angles.\\

\subsection{L-system}\label{lsystemmethodology}
In section \ref{design}, the L-system algorithm node was introduced in the preview A-Frame of the front-end design. This section discusses the details of the algorithm, how L-system models are generated and how parameters extracted in the analysis are interpreted by the algorithm. 

\subsubsection{L-system String Generation Algorithm}\label{lsystemgrammaralg}
To understand how Algorithm \ref{lsystemalg} works, a formal definition of the L-system and a set of L-system variables are needed.
The L-system can be defined by the triplet G = (V, $\omega$, P). Here:
\begin{itemize}
\item V consists of the variables I, B and a constant D. I is a branch segment, B is a branch segment candidate. D is a dead branch, which is a branch segment candidate that will not expand any further.
\item $\omega$ is the axiom. For example, the axiom could be III[/\&B][/\&B]. The number of I variables can be chosen.
\item P is the set of production rules, these are defined within the algorithm. For each variable, a new set of symbols and variables is returned.
\end{itemize}

Variables I and B share the following parameters:
\begin{itemize}
\item current iteration $\in \mathbb{N}_{\geq0}$ keeps track of the current iteration in the algorithm.
\item centerline index $\in \mathbb{Z}$ determines the centerline, with a unique centerline ID.
\end{itemize}

Parameters unique to I are:
\begin{itemize}
\item Radius $\in \mathbb{R}_{\geq0}$ of the cylinder geometry.
\item Length $\in \mathbb{R}_{\geq0}$ of the cylinder geometry.
\end{itemize}

Parameters unique to B are:
\begin{itemize}
\item Lifetime $\in \mathbb{Z}_{\geq0}$ of the variable.
\item Maximum centerline length $\in \mathbb{N}_{\geq0}$ determines the maximum lifetime of both I and B. If the centerline length exceeds this number, B will turn into a dead branch D. 
\end{itemize}

Furthermore, L-system parameters, italicized in Algorithm \ref{lsystemalg}, are defined as follows:
\begin{itemize}
\item stemLength $\in \mathbb{N}_{\geq0}$ defines the number of branch segments in the stem. This determines the number of I variables in the axiom.
\item radialIncrement $\in \mathbb{R}$ determines how much the thickness of each I grows with each iteration.
\item upwardsAngle, upwardsAngleSpread are the mean and the standard deviation of the normal distribution $\in [-180, 180]$ of upwards growing branch segments I.
\item lineLength, lineLengthStd $\in \mathbb{R}_{\geq0}$ are part of the cylinder geometry length distribution of I. 
\item nb\_axes\_prob is the probability a bifurcation will occur each time a new segment is built.
\item iterations, iterationsStd $\in \mathbb{N}_{\geq0}$ sets the maximum centerline length of B.
\item bifurcationAngle, bifurcationAngleSpread are the mean and the standard deviation of the bifurcation angle with a normal distribution $\in[-180, 180]$. 
\item maxOrder $\in \mathbb{N}_{\geq0}$ determines how spread out the branches will be from the stem. A higher maxOrder will ensure a longer lifetime for B variables at lower branch orders.
\end{itemize}

From these definitions it can be concluded that Algorithm \ref{lsystemalg} builds the scaffolding of the L-system with B variables and replaces B variables with I variables. I Variables are constant in the sense that they do not change to other symbols, only the segment radius changes with each iteration. The lifetime of B ensures that not many bifurcations occur at the start of the generation, which reduces intersections between segments in later generation stages. Besides their lifetime, B variables have a certain chance to create a bifurcation or turn into a dead branch with each iteration. As the variable and parameter definitions suggest, among other things, the rates at which bifurcations occur, lifetimes, number of iterations and radius increments can be preset and adjusted by the user. 

\begin{algorithm}
\caption{Pseudocode for L-system}\label{lsystemalg}
\begin{algorithmic}[1]
\State{Initialize stem with \textit{stemLength} branch segments and the first bifurcation}
\For{number of iterations}
    \For{each I}
        \State{Increment radius by \textit{radialIncrement}}
        \State{\Return{I}}
    \EndFor
    \For{each B}
        \If{max centerline length $<$ current iteration}
            \State{\Return{D}}
        \EndIf
        \If{time is shorter than maximum duration of B}
            \State{Increment time by 1 and iteration by 1}
            \State{$ \verb|^| \gets$ angle \textit{upwardsAngle} with standard deviation \textit{upwardsAngleSpread}}
            \State{I $\gets$ length \textit{lineLength} with standard deviation \textit{lineLengthStd}, radius \textit{radius}, \newline 
             \hspace*{6.2em} store iteration and centerline index}
            \State{\Return{$\verb|^|$IB}}
        \Else
            \State{Set time to 0, increment iteration and order by 1}
            \If{random chance of branching $>$ \textit{nb\_axes\_prob} \textbf{or} bifurcation has dead branch}
                \State{/ $\gets$ angle with random integer between -60 and 60}
                 \State{\Return{[/B]}}
            \Else
                \State{branches $\gets$ new array}
                \For{each branch in bifurcation}
                    \State{set max centerline length to \textit{iterations} with standard deviation \textit{iterationsStd}}
                    \State{set centerline index to random integer}
                    \State{/ $\gets$ angle with random integer between -60 and 60}
                    \State{\& $\gets$ angle \textit{bifurcationAngle} with standard deviation \textit{bifurcationAngleSpread} \newline
                    \hspace*{9.8em} and iteration correction}
                    \State{push [/\&B] to branches}
                \EndFor
                
                \State{\Return{branches}}
            \EndIf
        \EndIf
    \EndFor
\EndFor
\end{algorithmic}
\end{algorithm}

After generating the L-system object string, an interpreter is designed to create a 3D model corresponding to this object string using cylinder geometries. Before the geometries are pushed, the L-system model is checked for intersecting segments and pruning them if needed.\\

\subsubsection{Parameter Extraction from Model Analysis}\label{parextr}
Except the number of bifurcations, the centerline points, branch and centerline curvature, tortuosity and torsion, all parameters extracted in section \ref{modelanalysis} are used in the L-system algorithm as shown in table \ref{lsystemparameters}. First, the Bifurcation probability is set to $100 \%$ that means bifurcation is always happening. Furthermore, max. order has not been included in the analysis, so it is set to 0. Next, the mean number of iterations is set, and it is multiplied by two because it takes the algorithm at least two iterations to build a segment. The mean iterations cannot exceed 50, as the L-system model would grow too large. Subsequently, the upwards angle is imported, which is not allowed to be larger than 30 degrees or smaller than -30 degrees based on empirical experience. After that, the bifurcation angle mean is divided by two for each segment in the bifurcation. After importing the parameters and saving them on the client side, the L-system generation can commence.

\begin{table}[H]
    \begin{center}
        \begin{tabular}{||l||}
        \hline
        Parameter name (units) \\
        \hline
        \hline
        Segment radius (pixels) \\
        \hline
        Radial increment (pixels)  \\
        \hline
        Bifurcation probability \\
        \hline
        Radial segments \\
        \hline
        Iterations \\
        \hline
        Iterations std. \\
        \hline
        Max. order \\
        \hline
        Segment branch length (pixels) \\
        \hline
        Segment branch length std. (pixels) \\
        \hline
        Upwards angle (degrees) \\
        \hline
        Upwards angle std. (degrees) \\
        \hline
        Bifurcation angle (degrees) \\
        \hline
        Bifurcation angle std. (degrees) \\
        \hline
        \end{tabular}
    \end{center}
    \caption{A list of all available L-system parameters used in algorithm \ref{lsystemalg}.}\label{lsystemparameters}
\end{table}

\section{Results}\label{results}

\subsection{L-system Generation from Arborized 3D Models}\label{genfromarborized}
\subsubsection{lactiferous duct}\label{genfromarborizedmammary}
\begin{figure}[H]
     \centering
     \begin{subfigure}[b]{0.2\textwidth}
         \centering
         \includegraphics[height=3.5cm, angle=-90]{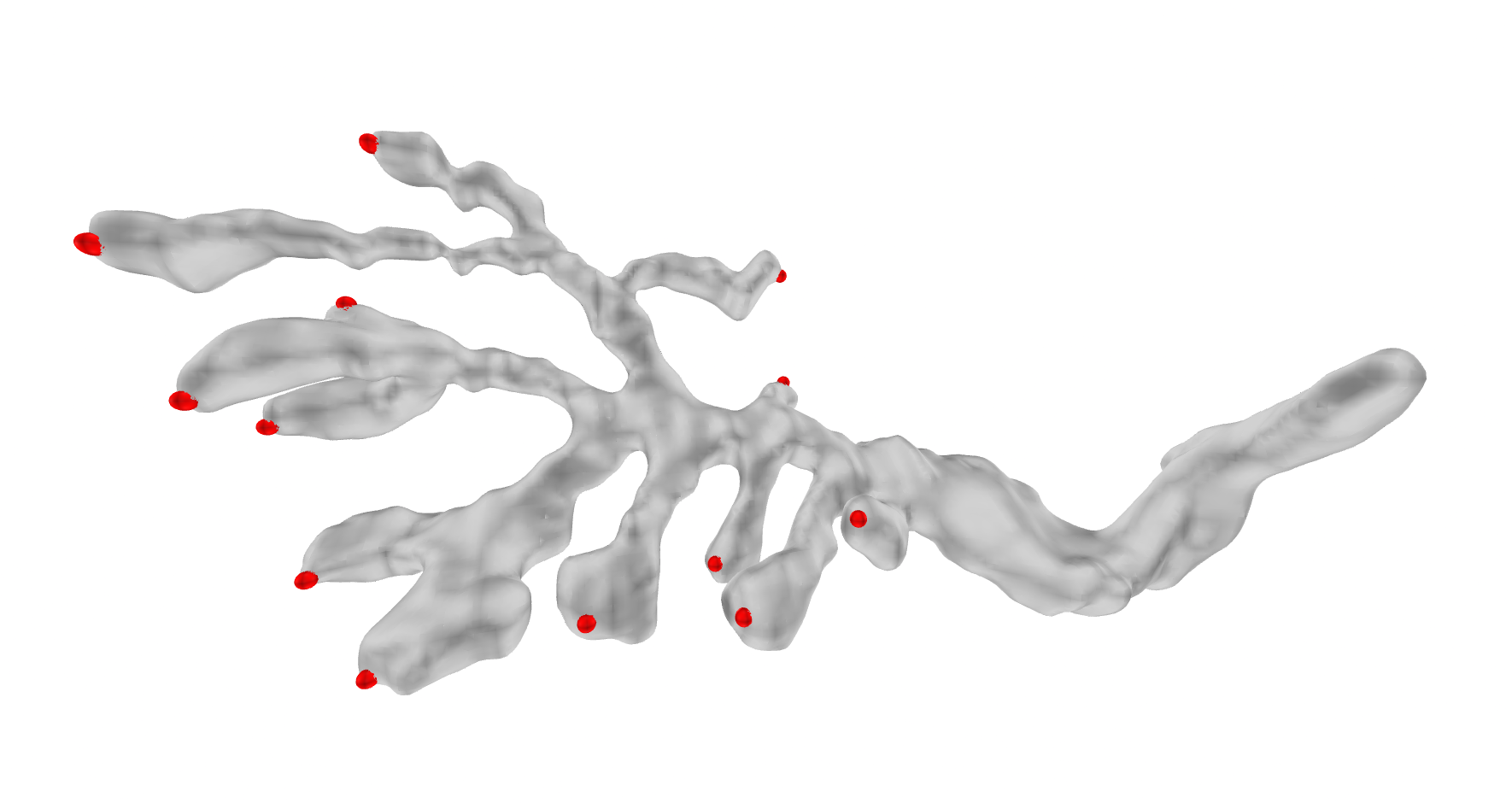}
         \caption{}
         \label{mammaryselected}
     \end{subfigure}
     \hfill
     \begin{subfigure}[b]{0.2\textwidth}
         \centering
         \includegraphics[height=3.5cm, angle=-90]{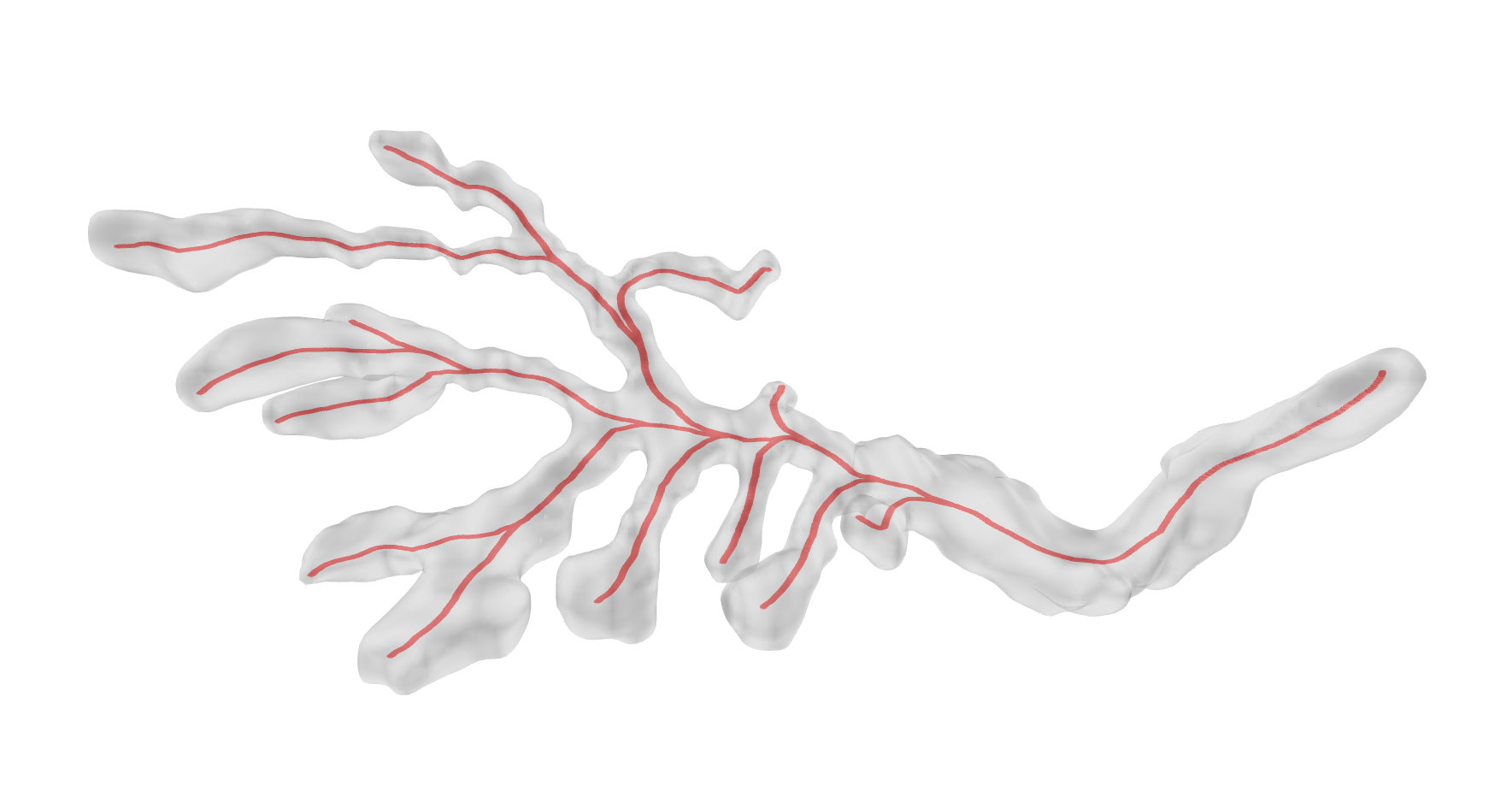}
         \caption{}
         \label{mammarycenterlinemodel}
     \end{subfigure}
     \hfill
        \caption{Model mesh of a lactiferous duct from newborn mice. (a) shows the model mesh and the endpoint selections used for the centerline creation (in red). (b) is the computed centerline model illustrated inside of the original mesh.}
        \label{mammarymodels}
\end{figure}

For the first case study, the section images of the lactiferous duct are acquired and a stack of aligned images is used as input for the optimized reconstruction pipeline \cite{Cao2012,10.1007/978-3-319-01781-5_16} of the 3D model as shown in figure \ref{mammarymodels}. The tree-like structure, starting at the thicker bottom stem and splitting to branches that vary in thickness, contains few bifurcations and consequentially centerlines. For this reason, it was used to test and verify the analysis and the L-system parameter extraction methods during the development of the web application. In this section, the results of the final analysis and L-system generation from the extracted parameters are found.\\

For the preparation of the analysis, endpoints are manually selected. In figure \ref{mammaryselected}, the selected endpoints are shown in red. The stem at the bottom was picked as the source point of the centerline structure. Endpoints are chosen based on distinct branches and where they approximately end in the mesh structure. Using this method, 13 endpoints were selected. After selecting the endpoints, the centerline in figure \ref{mammarycenterlinemodel} was extracted.\\

\begin{table}[H]
    \begin{center}
        \begin{tabular}{||l|l|l||}
        \hline
        Attribute name (units) & Mean & Std. \\
        \hline
        \hline
        Branch length (pixels) & 63.96 & 39.20 \\
        \hline
        Point radius (pixels) & 14.46 & 4.13 \\
        \hline
        Centerline length (pixels) & 582.56 & 123.60 \\
        \hline
        Stem length (\# iterations) & 4.32 & \\
        \hline
        Bifurcations & 11 & \\
        \hline
        Centerlines & 13 & \\
        \hline
        Branches & 24 & \\
        \hline
        Iterations & 4.79 & 1.93 \\
        \hline
        Bifurcation Angle (degrees)& 62.81 & 19.45 \\
        \hline
        \end{tabular}
    \end{center}
    \caption{Attributes extracted from the analysis of the model mesh in figure \ref{mammarymodels}. 
    Only the important attributes used for the L-system generation are shown in this table. Thus, attributes such as curvature, torsion and tortuosity are omitted.}
    \label{mammarydata}
\end{table}

\begin{figure}[H]
     \centering
     \begin{subfigure}[b]{0.2\textwidth}
         \centering
         \includegraphics[height=3cm]{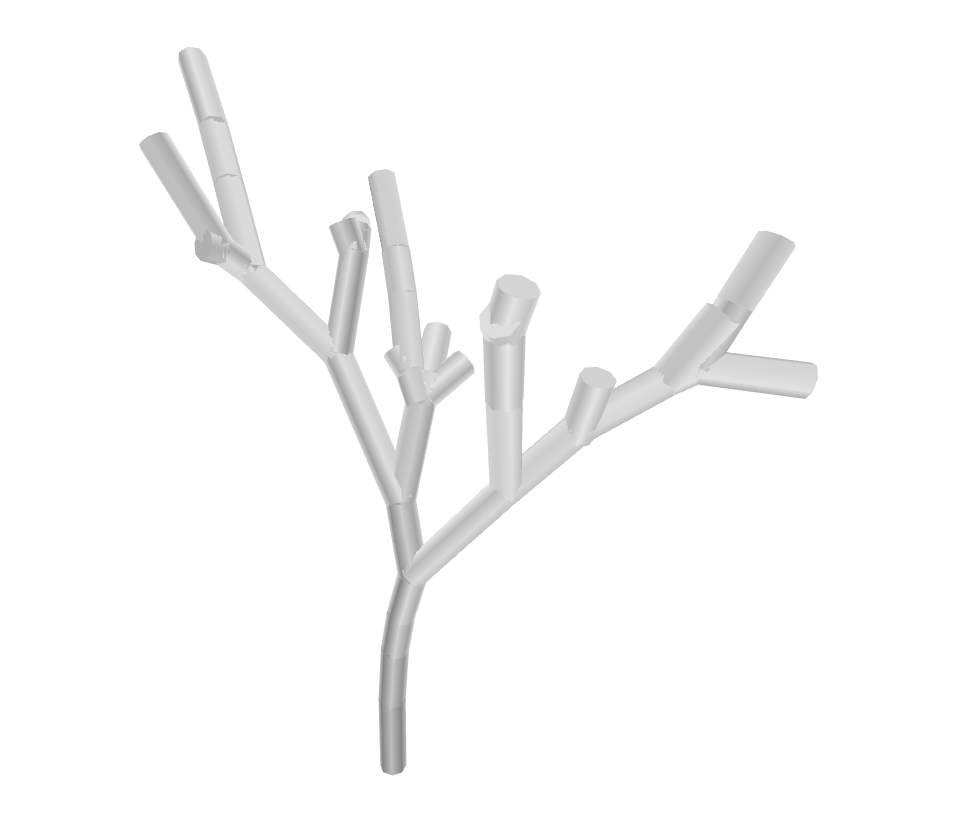}
         \caption{}
         \label{mammarylsystem}
     \end{subfigure}
     \hfill
     \begin{subfigure}[b]{0.2\textwidth}
         \centering
         \includegraphics[height=3cm]{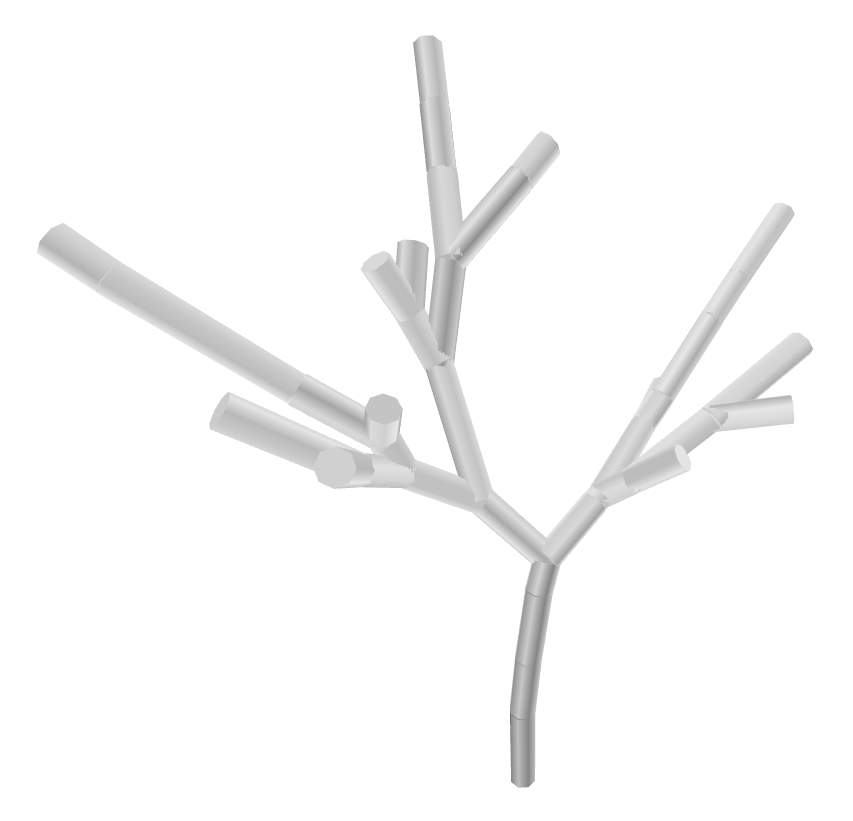}
         \caption{}
         \label{mammarylsystem2}
     \end{subfigure}
        \caption{A wide variety of randomly generated L-systems based on the data from table \ref{mammarydata} and with the number of iterations set to 10. The upwards curve of the stem was set to the standard value of 5 with a standard deviation of 2. (a) is an L-system with 16 centerlines and 43 branches. (b) is an L-system with 13 centerlines and 37 branches.}
        \label{mammarylsystems1}
\end{figure}

The data collected from the analysis both from the model and the centerline can be seen in table \ref{mammarydata}. A few observations can be made from the table. Firstly, from the 13 endpoints selected in figure \ref{mammaryselected}, 13 centerlines have been created, meaning there are no incorrectly analyzed centerlines. Secondly, the number of bifurcations does not entirely match the number of bifurcations which can be found in figure \ref{mammarycenterlinemodel}, because there is also one trifurcation mistaken for a bifurcation. Another observation is that the total number of iterations, which is in terms of how many mean branch lengths it takes to build one mean centerline, is approximately $9 \pm 2$. Finally, most of the varying data has a relatively high standard deviation, but the branch length, excluding the stem length, has the highest standard deviation. Figure \ref{mammarycenterlinemodel} also displays this large variety in branch lengths between bifurcations.\\
The parameters from table \ref{mammarydata} were then used to randomly generate L-systems. First, two L-systems were selected from many similarly generated systems using the given parameters, which can be found in figure \ref{mammarylsystems1}. During this process, some L-systems were created with colliding branches due to randomness of the algorithm. These results were left out by using the prune intersections option, as the original model does not contain intersections. Furthermore, models with similar data extracted from the original mesh were selected for the best representation. The three models were all generated with ten iterations, because building each segment needs at least two iterations. Additionally, they vary in branch lengths, number of centerlines and number of branches. The branch lengths range from frequent short branches in figure \ref{mammarylsystem} to frequent long branches or a mixture in figure \ref{mammarylsystem2}.

\begin{figure}[H]
     \centering
     \begin{subfigure}[b]{0.2\textwidth}
         \centering
         \includegraphics[height=3cm]{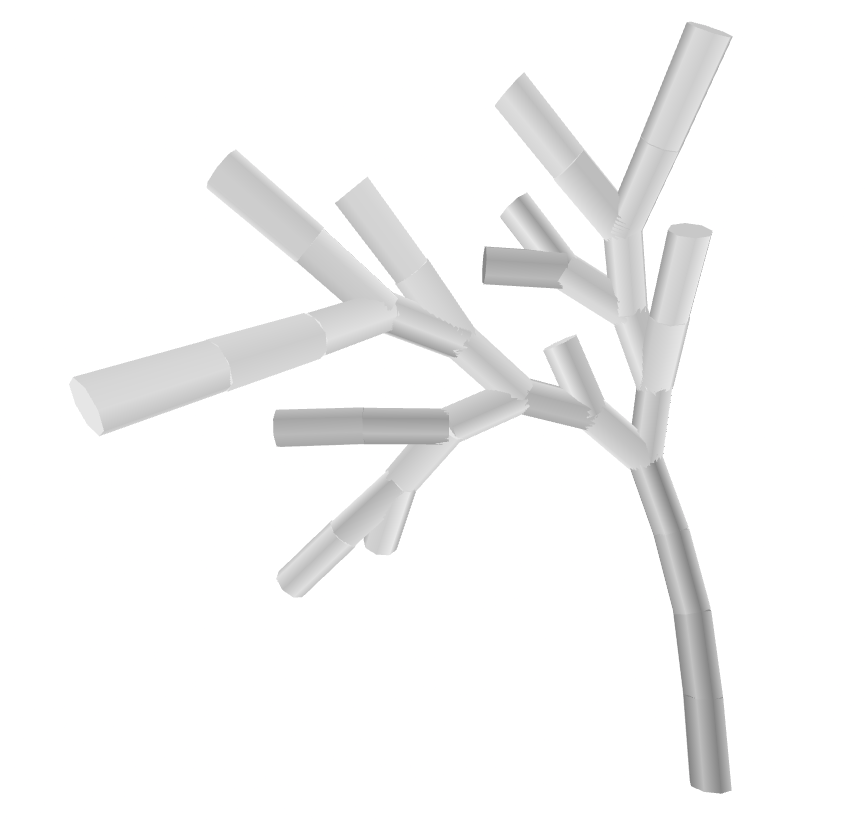}
         \caption{}
         \label{mammarylsystem3}
     \end{subfigure}
     \hfill
     \begin{subfigure}[b]{0.2\textwidth}
         \centering
         \includegraphics[height=3cm]{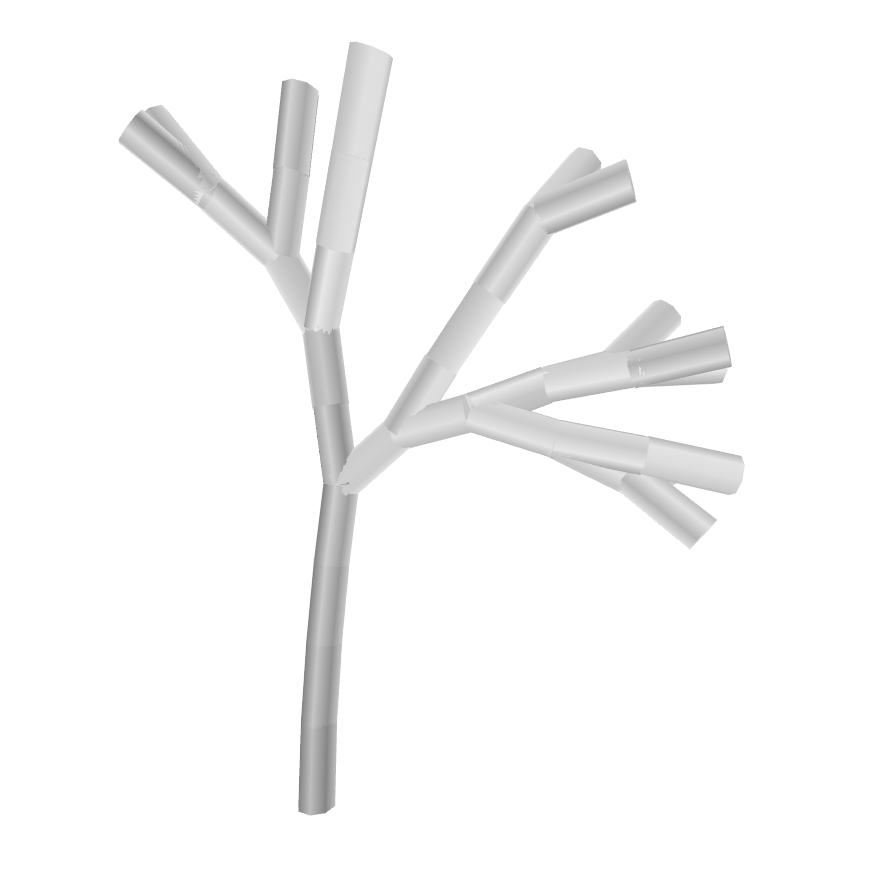}
         \caption{}
         \label{mammarylsystem5}
     \end{subfigure}
        \caption{Example of L-systems, based on the data from table \ref{mammarydata}. The upwards curve of the stem was set to the standard value of 5 with a standard deviation of 2. (a) is an L-system with 38 branches, 10 iterations and no line std. (b) is an L-system with 35 branches, 10 iterations, no iteration std., no branch length std. and 66\% branch probability.}
        \label{mammarylsystems2}
\end{figure}
Lastly, Figure \ref{mammarylsystems2} shows two L-systems with 13 centerlines and varying iterations, iteration standard deviation and no branch length standard deviation. These models were again selected from a pool of randomly generated L-systems. Figure \ref{mammarylsystem3} shows a model where the branch length standard deviation has been removed, which removes extra long branches since all the branch segment lengths are equal. The iteration standard deviation has been reduced by one iteration. Consequently, the variety in centerline lengths reduces. In Figure \ref{mammarylsystem5}, the variety in centerline lengths has been completely removed and the branch probability has been reduced to retain a diverse amount of centerlines and bifurcations. Due to the constant branch lengths, the overall shape of the L-systems in figure \ref{mammarylsystems2} is more compact compared to figure \ref{mammarylsystems1}.

\subsubsection{Artery}\label{genfromarborizedartery}
\begin{figure}[H]
     \centering
     \includegraphics[height=4cm]{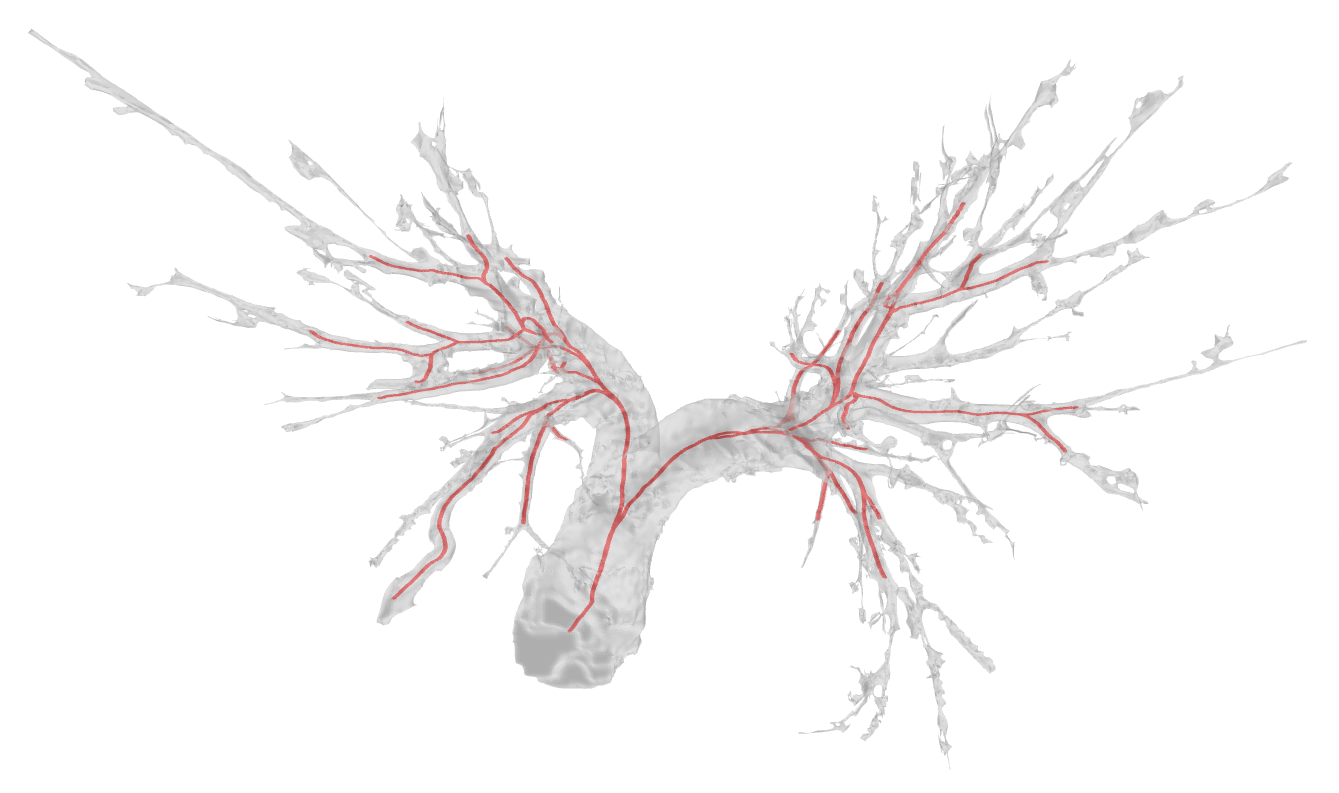}
     \caption{The original artery model with computed centerline.}
     \label{Arterycenterlinenotbroken}
\end{figure}

For the second case study, an artery 3D model was used. The source of the model is derived from \url{www.sketchfab.com}. This model has a more complex tree-like structure starting at the much thicker bottom stem and quickly decreasing in thickness. Some branches are hardly connected to each other with thin segments. Furthermore, the structure has a high branching order and therefore contains many bifurcations and centerlines. Thus, it is used to test the analysis and the L-system parameter extraction methods on a more complex structure. For the analysis, 40 endpoints were chosen. Figure \ref{Arterycenterlinenotbroken} shows the resulting centerline model in the original mesh. We observed that it contains many long branches compared to the lactiferous duct model.

The data collected both from the model and the centerline, can be found in table \ref{arterydata}. Once again, some observations can be made from this data. First, the number of centerlines is inline with the number of endpoints. Second, the number of bifurcations is correct since there are no trifurcations in this centerline model. Next, the total number of iterations is approximately $9 \pm 2$. Furthermore, most of the data has a relatively high standard deviation just like in table \ref{mammarydata}, even though there is more data available. Relatively, the bifurcation angle standard deviation contains the highest and the point radius the second highest standard deviation. Contrary to the bifurcation angle standard deviation, the relatively large branch length and point radius standard deviations can be observed by the alternation between long and short branches together with the thick inner branches close to the stem and thin outer branches further away from the stem.

\begin{table}[H]
    \begin{center}
        \begin{tabular}{||l|l|l||}
        \hline
        Attribute name (units) & Mean & Std. \\
        \hline
        \hline
        Branch length (pixels) & 17.25 & 12.70 \\
        \hline
        Point radius (pixels) & 6.39 & 4.48 \\
        \hline
        Centerline length (pixels) & 150.04 & 20.86 \\
        \hline
        Stem Length (\# iterations) & 2 & \\
        \hline
        Bifurcations & 32 & \\
        \hline
        Centerlines & 40 & \\
        \hline
        Branches & 72 & \\
        \hline
        Iterations & 6.82 & 1.20 \\
        \hline
        Bifurcation angle (degrees)& 44.40 & 37.68 \\
        \hline
        \end{tabular}
    \end{center}
    \caption{Attributes extracted from the analysis of the artery model. In the first column, the attribute name and units are shown. The second and third column show the mean and the standard deviation if applicable.}\label{arterydata}
\end{table}

\begin{figure}[H]
     \centering
     \begin{subfigure}[b]{0.2\textwidth}
         \centering
         \includegraphics[height=2.5cm]{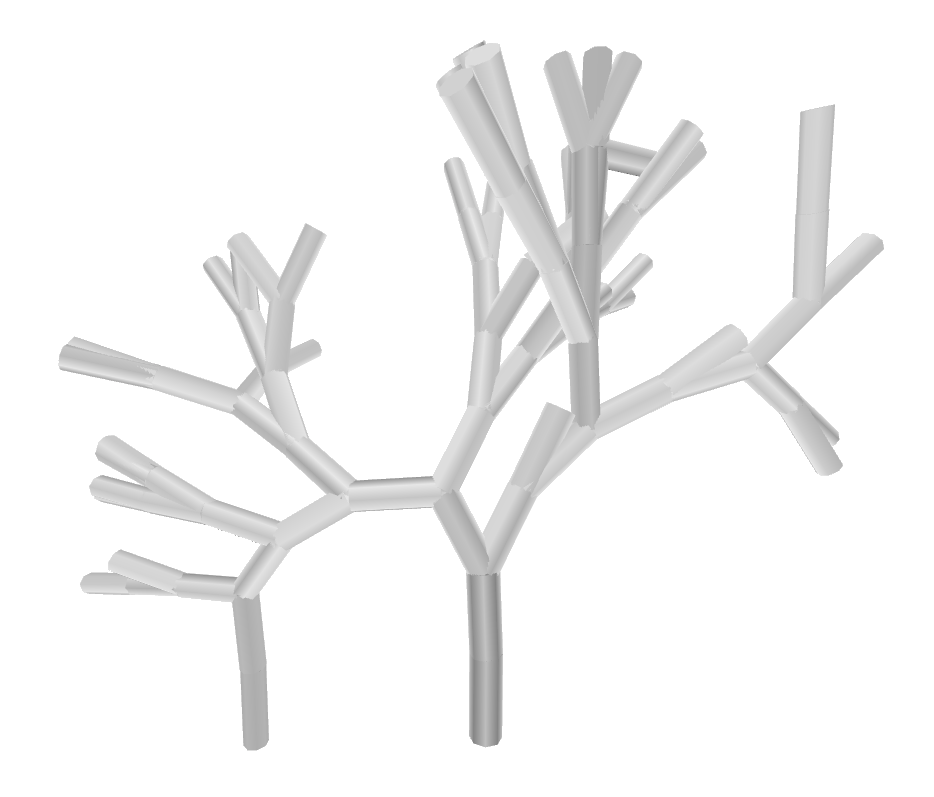}
         \caption{}
         \label{arterylsystem4}
     \end{subfigure}
     \hfill
     \begin{subfigure}[b]{0.2\textwidth}
         \centering
         \includegraphics[height=2.5cm]{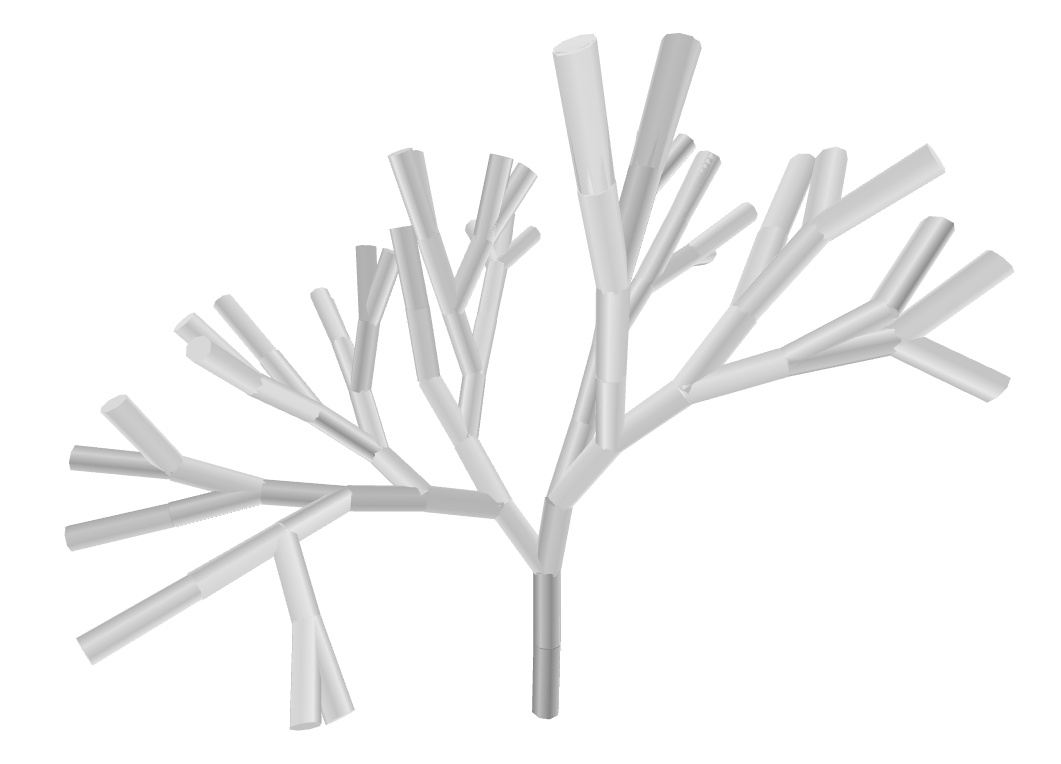}
         \caption{}
         \label{arterylsystem5}
     \end{subfigure}
        \caption{Two randomly generated L-systems based on the data from table \ref{arterydata}. For all of the systems, the number of iterations was set to 12. The upwards curve of the stem was set to the standard value of 5 with a standard deviation of 2. (a) is an L-system with 40 centerlines, 85 branches, 2.39 segment radius, no branch length std. and one iteration std. (b) is an L-system with 38 centerlines, 86 branches, 2.39 segment radius, no branch length std, no iteration std. and 76\% branch probability.}
        \label{arterylsystems2}
\end{figure}

For the L-system generation from the attributes acquired in table \ref{arterydata}, We observed an abundance of colliding branches and centerlines, indicating that with the given 14 iterations, too many branches were generated. Figure \ref{arterylsystems2} shows L-systems without the branch length standard deviation. In Figure \ref{arterylsystem4}, the iteration standard deviation was reduced, which is demonstrated by the reduced number of outliers and radial development. Figure \ref{arterylsystem5} has no iteration and branch length standard deviations but a branch probability. This lead to equal centerline lengths and no more cut-off centerlines compared to the systems generated before.

Lastly, what can be observed from all the L-systems which have been generated from the lactiferous duct and artery models, is the increased number of branches compared to the number of centerlines. The data in table \ref{mammarydata} and \ref{arterydata} show that the original models require less branches to obtain the same number of centerlines compared to what is produced by the algorithm.

\section{Conclusion and Discussion}\label{conclusiondiscussion}
In this paper, we have focused on combining visualization, analysis and L-system abstraction in a web-environment. We generalized this functionality for numerous arborized 3D biological structures. 
The visualizing capabilities were shown to be similar with other visualization tools such as MeshLab~\cite{meshlab}. On the other hand, the ability to visualize 3D models is limited by the user's (web) resources, which restricts the complexity of the models that can be rendered.\\
Secondly, it has been illustrated that the functionality is sufficient to perform proper statistical analysis on branch-like structures and extract useful L-system parameters. An important condition for the analysis method is a sufficient branch radius. However, no specification was given on which radii satisfy these conditions. Another important condition, stated in the VMTK documentation, requires the 3D model mesh to be close-ended. \\
Next, L-system generation from extracted parameters was shown. Changing the upwards and bifurcation angle standard deviation did not seem to result in much of a difference compared to the related work it was derived from~\cite{verbeek2020systems}. In contrast, the iteration and especially the branch length standard deviation did have a significant impact on the resulting models. Removing the branch length standard deviation made the overall shape look less chaotic and more comparable to the centerline models in both examples. Depending on the type of abstraction, either the original branch probability or the iteration standard deviation can be utilized to add more variety to centerline lengths. Thus, extracted parameters are viable to generate L-systems, particularly branch lengths, radii, angles and iteration standard deviation.\\
From these key results, we can conclude the method described in this paper could be applied to develop a web application that is in fact capable of visualizing, analyzing and creating L-system abstractions of arborized 3D models. In addition, easy accessibility and portability can provide workflow-improving benefits and the applied web development methods contribute to extendibility.

For the future work, concerning visualization, larger and more complex models could be supported with server side rendering~\cite{Sawicki2013} or by rendering only specific sections of the 3D model. The benefit of this is that the client requires less resources, but at the cost of more server resources and the loss of real-time navigation and interaction. Besides visualization, more (arborized) 3D model analysis functionality and methods can be further extended. This could include automated endpoint selection found in Slicer~\cite{Kikinis2014}, model pre-processing methods for open-ended model analysis and error reduction~\cite{VMTK} or support for analysis of multiple similar 3D models at once.\\
For L-system generation, additional statistical research can be conducted on the distributions of the parameters and how the data can be applied optimally, particularly when dealing with relatively high standard deviations. This also includes the right estimate of iterations, which consistently failed during the testing phase. Further experiments on L-system parameters and resulting abstractions may give rise to models which are more consistent and in accordance with the requirements.\\
To conclude, VR features that are readily available within the A-Frame framework can be utilized. This includes VR interaction methods with the user interface and new visualization capabilities (such as a free-floating camera) that can change the way 3D structures are analyzed and perceived.

\bibliographystyle{apalike}
{\small
\bibliography{bibliography}}

\end{document}